\documentclass[aps,pre,superscriptaddress,twocolumn,amsmath,amssymb,showpacs]{revtex4}
\usepackage{graphicx}
\usepackage{dcolumn}
\usepackage{bm}
\newcommand{\ignore}[1]{}
\usepackage{epsfig}
\usepackage{color}

\begin{document}

\title{Vegetation pattern formation in semiarid systems without facilitative mechanisms}

\author{Ricardo Mart\'inez-Garc\'ia}
\affiliation{IFISC, Instituto de F\'isica Interdisciplinar y Sistemas Complejos (CSIC-UIB),
 E-07122 Palma de Mallorca, Spain}

\author{Justin M. Calabrese}
\affiliation{Conservation Ecology Center, Smithsonian Conservation Biology Institute, National Zoological Park,
1500 Remount Rd., Front Royal, VA 22630, USA.} 

\author{Emilio Hernandez Garcia} 
\affiliation{IFISC, Instituto de F\'isica Interdisciplinar y Sistemas Complejos (CSIC-UIB),
 E-07122 Palma de Mallorca, Spain}

\author{Crist\'obal L\'opez} 
\affiliation{IFISC, Instituto de F\'isica Interdisciplinar y Sistemas Complejos (CSIC-UIB),
 E-07122 Palma de Mallorca, Spain}

\begin{abstract}

Regular vegetation patterns in semiarid ecosystems are believed to arise
from the interplay between long-range competition and facilitation processes
acting at smaller distances. We show that, under rather general
conditions, long-range competition alone may be enough to shape these
patterns. To this end we propose a simple, general model for the dynamics of
vegetation, which includes only long-range competition between plants. Competition is introduced
through a nonlocal term, where the kernel function quantifies the
intensity of the interaction. We recover the full
spectrum of spatial structures typical of vegetation models
that also account for facilitation in addition to competition.

\end{abstract}

\maketitle

\section{Introduction}
\label{sec:Introduction} Regular patterns  and spatial
organization of vegetation have been observed in many arid and
semiarid ecosystems worldwide, covering a diverse range of
plant taxa and soil types \citep{Klausmeier,
RietkerkTrends,thompson}. A key common ingredient in these
systems is that plant growth is severely limited by water
availability, and thus plants likely compete strongly for water
\citep{TheAmNat2002}. The study of such patterns is especially
interesting because their features may reveal much about the
underlying physical and biological processes that generated
them in addition to giving information on the characteristics of
the ecosystem. It is possible, for instance, to infer their
resilience against anthropogenic disturbances or climatic
changes that could cause abrupt shifts in the system and lead
it to a desert state
\citep{vandekoppel2002,D'Odoricoa,D'Odoricob}.

Much research has therefore focused on identifying the mechanisms
that can produce spatial patterning in water limited systems
\citep{LefeverLejeune,Klausmeier,Pueyo}. An important class of
deterministic vegetation models (i.e., those not considering
noise associated with random disturbances) that can produce
regular patterns are the kernel-based models
\citep{D'Odoricoa}. These models produce patterns via a
symmetry-breaking instability (i.e., a mechanism by which the
symmetric-homogeneous state loses stability and a periodic
pattern is created) that has its origins in the interplay
between short-range facilitation and long-range competition
\citep{D'Odoricob,RietkerkTrends,BorgognoRG}, with field
observations confirming this hypothesis in some landscapes
\citep{Dunkerley}. Therefore it has been long assumed that both
of these mechanisms must be present in semi-arid systems to
account for observed vegetation patterns, although quantifying
the importance of each one has proven to be a difficult and
contentious task \citep{Barbier,Veblen}. A key role theory can
play here is to identify the minimal requirements for pattern
formation to occur. \citep{RietkerkTrends} have speculated that
pattern formation, under certain conditions, could occur
without short-range facilitation. More recently, a model
proposed for mesic savannas included fire and plant-plant
competition as key ingredients \citep{MartinezGarciaJTB}. Fire
introduces a positive feedback so that this model considers
both competition and facilitation mechanisms. However, the
model still produced regular patterns even when the
facilitative interaction, fire, was considered at its very
short-range (in fact, local) limit. These considerations
suggest that local facilitation may be superfluous for pattern
formation, and that a deeper exploration of the range of
conditions under which pattern formation can occur in the
absence of facilitation is therefore warranted.

Here, we study a simple but quite general single-variable model
that considers the time evolution of vegetation density in
water-limited regions, with only competitive interactions among
plants. We show that when only a single broadly applicable
condition is met, that competitive interactions have a finite
range, the full set of regular patterns formerly attributed to
the interaction between short-range facilitation and
long-distance competition can be produced in the absence of
facilitation.

\section{The Model}
\label{sec:Model}

Arid and semiarid ecosystems are typified by patches of
vegetation interspersed with bare ground. Water is a very
limited resource for which juvenile plants must compete with
those that have already established. Logistic-type
population models have been used in a wide variety of
applications including semiarid systems and savannas
\citep{calabrese}, and thus form a reasonable and very general
starting point. Specifically, we consider the
large-scale long-time description of the model in terms of a
continuous-time evolution equation for the density of trees,
$\rho ({\bf x},t)$. Death occurs at a constant rate $\alpha$,
whereas population growth occurs via a sequence of seed
production, dispersal, and seed establishment processes. Seed
production occurs at a rate $\beta_0$ per plant. For simplicity
we consider dispersal to be purely local and then if all seeds
would give rise to new  plants the growth rate would be
$\beta_0\rho({\bf x},t)$. But once a seed lands, it will have
to overcome competition in order to establish as a
new plant. We consider two different competition mechanisms:
First, space availability alone limits density to a maximum
value given by $\rho_{max}$. Thus, $0\leq \rho ({\bf x},t) \leq
\rho_{max}$. The proportion of available space at site ${\bf
x}$ is $1-\rho({\bf x},t)/\rho_{max}$, so that the growth rate
given by seed production should be reduced by this factor.
Second, once the seed germinates, it has to overcome
competition for resources with other plants. This is included in the model by an additional factor
$r=r(\tilde{\rho},\delta)$, $0\leq r \leq 1$, which is the
probability of overcoming competition. This probability decreases with increasing average vegetation density within a
neighborhood $\tilde \rho$, and the strength of this decrease depends on the competition intensity parameter, $\delta$. Higher values of $\delta$
represent more arid lands, and thus stronger competition for
water. In the following, we measure density in units so that
$\rho_{max}=1$. Combining all processes, the evolution equation
for the density then takes the form:

\begin{equation} \label{model}
\frac{\partial \rho({\bf x},t)}{\partial t}=\beta_0 r(\tilde{\rho},\delta)\rho({\bf x},t)(1-\rho({\bf x},t))-\alpha\rho({\bf x},t).
\end{equation}

$\tilde{\rho}=\tilde{\rho}({\bf x},t)$ is the nonlocal density
of vegetation that is obtained by averaging (with a proper
weighting function) the density of plants in a neighborhood:

\begin{equation}\label{nonlocal}
\tilde{\rho}({\bf x},t)=\int G(|{\bf x}-{\bf x'}|)\rho({\bf x'},t)d{\bf x'},
\end{equation}
 where $G({\bf x})$ is a normalized kernel function, which
accounts for the weighted mean vegetation density, and defines
the neighborhood of the plant. A Laplacian term could be
included in the r.h.s of Eq. (\ref{model}) as a way to model long range
seed dispersal, but doing so would not qualitatively change our
results, so we have left it out.

In previous kernel-based vegetation models \citep{LefeverLejeune,D'Odoricoa},
the kernel function contained
information on the class of interactions present in the system,
that were both competitive (inhibitory) and facilitative. On
the contrary, we introduce purely competitive interactions
through the nonlocal function $r(\tilde{\rho},\delta)$, where
the kernel defines the area of influence of a focal plant, and
how its influence decays with distance. Competition is included
by assuming that the probability of establishment
$r$ decreases with increasing vegetation density in the
surroundings:

\begin{equation}
\frac{\partial r(\tilde{\rho},\delta)}{\partial\tilde{\rho}}\leq 0. \label{cond2}
\end{equation}
As $\delta$ modulates the strength of the competition, it must
be that $r(\tilde{\rho},\delta=0)=1$, and
$r(\tilde{\rho},\delta\rightarrow\infty)=0$. This means that
when water is abundant ($\delta=0$) competition for water is
not important  ($r=1$), whereas new plants cannot establish in
the limit of extremely arid systems, $\delta\rightarrow\infty$.

Note the generality of the vegetation competition model: a
spatially nonlocal population growth term of
logistic type with rate fulfilling Eq. (\ref{cond2}), and a
linear death term.  We note that previous work has
shown that competitive interactions entering multiplicatively
in the death term \citep{Birch2006} or additively in the model
equation \citep{BorgognoRG} may also lead to pattern
formation. A complete description of our model should specify
both the kernel function $G$ as well as $r$, but we can go
further with the analysis in general terms.

\section{Results}

The possible homogenous stationary values of the density for
Equation (\ref{model}) are: a) no
vegetation $\rho=0$,  and b) the  vegetated state $\rho=\rho_{0}$.
The system will show
either one or the other depending on the relationship between
the birth and death rates, $\beta_{0}$ and $\alpha$ \citep{calabrese}.
The non-trivial homogeneous stationary solution, $\rho_{0}$, can be obtained by solving

\begin{equation}\label{homo}
\beta_{0}r(\rho_{0},\delta)(1-\rho_{0})-\alpha=0,
\end{equation}
that has only one solution in the interval
$\rho_{0}\in[0,1]$ because of the conditions imposed on the
function $r$ in Eq.(\ref{cond2}). We now ask if this
stationary solution gives rise to periodic structures via a
symmetry-breaking instability as happens in other models that
include not only competition but also facilitation mechanisms
in the vegetation interactions \citep{BorgognoRG}. To explore
this possibility in our model, we perform a linear stability
analysis \citep{cross} by adding a small perturbation to the
stationary solution, so $\rho({\bf
x},t)=\rho_{0}+\epsilon\psi({\bf x},t)$, with $\epsilon\ll 1$.
Technical details of this derivation may be found in Appendix
\ref{deriv}. We obtain a perturbation growth rate

\begin{equation}
\label{condition}
 \lambda({\bf k})=-\alpha\rho_{0}\left[\frac{1}{1-\rho_{0}}-\frac{r'(\rho_{0},\delta)}{r(\rho_{0},\delta)}\hat{G}({\bf k})\right],
\end{equation}
where $\hat{G}({\bf k})$ is the Fourier transform of the
kernel, $\hat{G}({\bf k})=\int G({\bf x})\exp(i{\bf k}\cdot{\bf
x})d{\bf x}$, and
$r'(\rho_{0},\delta)\equiv\left(\frac{\partial r}{\partial
\tilde\rho}\right)_{\tilde\rho=\rho_{0}}$.

Patterns appear if the maximum of the growth rate (i.e., of the
most unstable mode), $\lambda(k_{c})$, is positive, which means
that the perturbation grows with time. From
Eq.~(\ref{condition}), this is only possible if the Fourier
transform of the kernel function, $\hat{G}({\bf k})$, takes
negative values, since  $r'(\rho_{0},\delta)<0$. This happens,
for example, for all stretched exponentials $G(|{\bf
x}|)\propto \exp{(-|{\bf x}/R|^p})$ with $p>2$, where $R$ is a
typical interaction length \citep{pigolotti,pigolotti2010}.
Kernels satisfying this criterion have broader shoulders and
shorter tails (i.e., are more platykurtic) than the Gaussian
function, which is obtained for $p=2$. In reality, any
competitive interaction among plants will have finite range
because their roots, which mediate the interaction, have finite
length. The interaction range $R$ between two plants
will be twice the typical root length. Kernels with finite range can,
in general, be modeled by considering a truncated function such
that $G(|x|)=CF(|x|)\Pi(|x|)$, where $C$ is a normalization
constant, $\Pi(x)$ is a unit-step function defined as
$\Pi(x)=1$ if $|x| \leq R$ and $\Pi(x)=0$ if $|x| > R$, and
$F(|x|)$ is a function of the distance that models the
interactions among the plants. Because of the finite range in
the kernel function, the Fourier transform will show
oscillations and thus will always take negative values. The
functional form of the probability of surviving the
competition, $r(\tilde{\rho},\delta)$, changes only the
parameter regime where patterns first develop, but they will
appear in the system, regardless of its form, for
$r'(\rho_{0},\delta)/r(\rho_{0},\delta)$ large enough.

For the rest of our analysis, we will use $F(x)=1$, so the
kernel is given by $G(x)=1/\pi R^2$ if $|x| \leq R$ and
$G(x)=0$ if $|x| > R$, which defines an interaction area of
radius $R$ (that is, roots of typical length
$R/2$). Its Fourier transform (in two dimensions) is

\begin{equation}\label{tophat}
 \hat{G}({\bf k})=\frac{2J_{1}(|{\bf k}|R)}{|{\bf k}|R},
\end{equation}
where $J_{1}(|{\bf k}|R)$ is the first-order Bessel function.
We will further specify the model by assuming
particular forms for the growth rates.
Let us consider a probability of surviving competition given by

\begin{equation}\label{pldecay}
 r(\tilde{\rho},\delta)=\frac{1}{(1+\delta\tilde{\rho})^{q}},
\end{equation}
with $q>0$. In the particular case of $q=1$, the homogeneous
density, $\rho_{0}$, and the perturbation growth rate,
$\lambda$, can be obtained analytically. Numerical evaluations
must be done if $q\neq1$. In the following, for simplicity, we
consider the case $q=1$ and only briefly discuss other values.
The nontrivial stationary solution, $\rho_{0}\neq0$, can be
obtained analytically

\begin{equation}\label{rhostat}
 \rho_{0}=\frac{\beta_{0}-\alpha}{\beta_{0}+\alpha\delta},
\end{equation}
where $\beta_{0}\geq\alpha$. Equation~(\ref{rhostat}) shows
that the homogeneous density of trees in the stationary state
decays as $\sim\delta^{-1}$ with increasing competition
strength (i.e., large $\delta$). It can be analytically shown
that the same dependence of $\rho_{0}$ on large $\delta$ occurs
for any value of $q$.

From Eq.~(\ref{condition}), the growth rate of
perturbations can also be calculated

  \begin{figure}
  \noindent\includegraphics[width=\columnwidth]{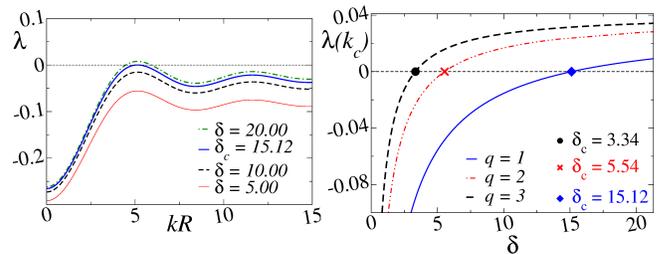}
  \caption{ (Left) Perturbation growth rate given by Eq.~(\ref{dispower}) using a
  unit-step kernel for different values of $\delta$.
  From bottom to top $\delta=5.00$, $\delta=10.00$, $\delta_{c}=15.12$, $\delta=20.00$. (Right) $\lambda(k_{c})$, as a function of $\delta$, using
  $r(\tilde{\rho},\delta)$ given by Eq.~(\ref{pldecay}). From right to left $q=1$, $q=2$, $q=3$. In both panels, other parameters: $\beta_{0}=1.0$ and $\alpha=0.5$.}
  \label{figpldisp}
  \end{figure}
  
\begin{equation}\label{dispower}
 \lambda({\bf k})=\frac{(\alpha-\beta_{0})(\beta_{0}+\alpha\delta\hat{G}({\bf k}))}{\beta_{0}(1+\delta)},
\end{equation}
and is shown in Figure \ref{figpldisp} (Left) for different
values of the competition strength. When the growth rate of the
most unstable mode (i.e. the maximum of $\lambda (k)$),
$k_{c}$, becomes positive, patterns emerge in the system
\citep{BorgognoRG}. To obtain the critical value of the
competition parameter at the transition to patterns,
$\delta_{c}$, we have to calculate the most unstable mode as
the first extreme of $\lambda (\bf k)$ at $k\neq0$, i.e., the
first zero of the derivative of $\hat{G}(\bf k)$. This value
only depends on $R$ (the range defining $G (\bf r)$) and it is
$k_{c}=5.136/R$. Because a periodic pattern of $n$ cells of
vegetation is characterized by a wavenumber $k_{c}=2\pi n/L$,
where $L$ is the system size, the typical distance between two
maxima of vegetation, $d=L/n$, is given by $d \approx 1.22R$.
This value changes depending on the kernel, but in the case of
kernels with a finite range (i.e. truncated by a unit step
function of radius $R$) it is always on this order.
The critical wavenumber is determined mainly by the
contribution of the unit step function to the Fourier
transform, which is always the same. This result is also
independent of the other parameters of the system, and shows
that the nonlocal competition mechanism is responsible for the
formation of patterns in the system.

To identify the parameter values for the transition to
patterns, we solve $\lambda({\bf k}_{c})=0$ in
Eq.~(\ref{dispower}), which shows that patterns emerge when
competition strength exceeds
$\delta_{c}=-\beta_{0}/\alpha\hat{G}(\mathbf{k_{c}})$, which is
positive because $\hat{G}(\mathbf{k_{c}})<0$. Figure
\ref{figpldisp} (Right) shows the growth rate of the most
unstable mode as a function of competition strength for
different values of the exponent $q$ for fixed values
$\beta_{0}=1$, and $\alpha=0.5$. Note that the critical value
of the competition parameter depends on the functional form of
$r$. This dependence could be used to tune the value of $q$ to
have a realistic competition strength for the transition to
patterns, provided that one has sufficient data.

\begin{figure}
  \noindent\includegraphics[width=0.5\columnwidth]{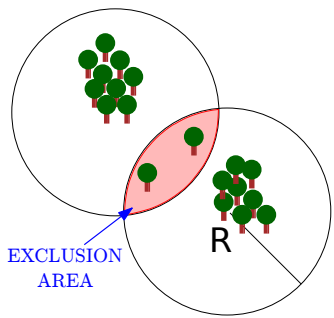}
  \caption{Schematic representation of the formation of exclusion areas, where plants have to compete with two different
  vegetation patches, whereas plants in each patch compete only with individuals in its own patch.}
  \label{exclusion}
  \end{figure}

We can also explain the separation length between clusters of
plants using ecological arguments. Consider a random and
inhomogeneous distribution of plants. Maxima of this
distribution identify places with the highest plant density.
Imagine that two such maxima occur at a distance larger than
$R$ but smaller than $2R$ from each other. There will be no
direct interaction between the roots of plants in these
different patches, because they are separated by a distance
larger than the interaction range $R$ (twice the
root extension). But there is an area in-between which is
simultaneously within the range of both patches. Compared with
plants occurring inside a cluster, which only have to compete
with plants in their own cluster, those that occur in-between
clusters will experience stronger competition and will
therefore tend to disappear (Figure \ref{exclusion}).
We call
these regions featuring very strong competition {\it exclusion
areas}, consistent with previous studies of competition-driven
spatial pattern formation
\citep{Hernandez-Garcia,pigolotti,pigolotti2010}. The
disappearance of plants in these exclusion areas in turn
reduces competition on the two well-populated patches, so that
a positive feedback appears reinforcing the establishment of
plants in patches periodically separated with a distance
between $R$ and $2R$. We stress again that competition alone is
responsible for the symmetry breaking instability, and no
facilitative interactions are needed for pattern formation.

\begin{figure}
  \noindent\includegraphics[width=\columnwidth]{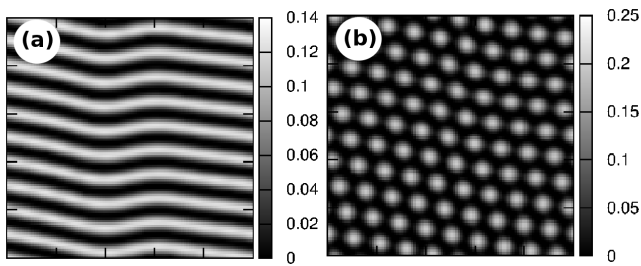}
  \caption{Steady spatial structures shown by the model using the $r(\tilde{\rho},\delta)$ given by Eq.~(\ref{pldecay})
  with $q=1$. Darker grey levels represent smaller densities. (a) Vegetation stripes, $\delta=16.0$.
  (b) Vegetation spots, $\delta=17.0$.  Other parameters: $\beta_{0}=1.0$ and $\alpha=0.5$}.
  \label{patterns}
  \end{figure}

Finally, we have numerically integrated Eq.~(\ref{model}) in a
patch of $10^{4}$~m$^{2}$ with periodic boundary conditions and
a competition range of $R=8$~m. Time stepping is done with an
Euler algorithm. The results (see Figure \ref{patterns})
exhibit steady striped and spotted vegetation patterns. This
spectrum of patterns, typical in pattern formation arising from
symmetry breaking, is also observed in models that include a
short-range facilitation mechanism in addition to long-range
competition \citep{LejeuneTlidi,TheAmNat2002}.

We have checked that similar results can be obtained for
different growth rates, for example stretched exponentials:

\begin{equation}\label{rexp}
r(\tilde{\rho},\delta)= e^{-\delta\tilde{\rho}^{p}}.
\end{equation}
This further confirms our result that competition is the only
necessary ingredient for the formation of vegetation patterns
in the present framework, and that this does not depend on the
functional form of the probability of surviving competition
(growth rate) provided it verifies the requirements given by
Eq. (\ref{cond2}).

\section{Summary and conclusions}\label{sec:summary}

We have studied the formation of spatial structures of
vegetation in arid and semiarid landscapes, where water is a
limiting resource for which plants must compete. We have
considered a simple model with a linear death and a
logistic-type growth term in which the growth/birth
rate depends on the average vegetation density in the
surroundings. Competition enters the model by inhibiting plant
growth when local density increases. Arid and semiarid
ecosystems correspond to intermediate and high values of the
model parameter $\delta$, that modulates competition intensity.
Our main result is that patterns appear in the
system despite the absence of short range facilitation
mechanisms, and that these patterns exist regardless of the
functional form of the nonlocal growth rate, provided that
competition is strong enough. Previous studies have included an
interaction term that accounts for a short-range positive
effects of high local vegetation density, as well as for
long-range competition. This combination of mechanisms is
justified by arguing that water percolates more readily through
the soil in vegetated areas \citep{D'Odorico2005}
(short-range), and that plants compete for water resources over
greater distances via long lateral roots (long-range). In
addition, recent studies on mesic savannas
\citep{MartinezGarciaJTB} have shown that in the
infinitesimally short limit (i.e., local) of facilitative
interactions, tree patterns still appear in the system.
In contrast with these studies, in the simple
situation that we present competition is the only mechanism
responsible for pattern formation, provided that the Fourier
transform of the kernel function takes negative values. It is
important to note that the simple requirement of
just competitive interactions among plants is rather general
and does not depend on the way these interactions are
introduced in the model. For example considering a
death term that increases with nonlocal density through a
competition kernel also gives rise to pattern formation (see
\citep{Birch2006} for a related study in a different context).
In addition, if nonlocal competition enters in the model
additively, one may also obtain spatial structures that are
determined by the properties of the Fourier transform of the
kernel.

The finite interaction range typical of any real competitive
interaction implies a truncation of the kernel function, and as
we have shown, this greatly expands the range of kernels that
can lead to pattern formation. The development of exclusion
zones between maxima of the plant density, where competition is
stronger, is the mechanism by which patterns emerge, because
competition tends to prevent the growth of vegetation in those
regions.

We have demonstrated that our vegetation model recovers the
gapped and striped patterns observed in arid and semiarid
landscapes when the finite range of the competitive interaction
is considered, and thus there is a kernel function whose
Fourier transform may have negative values. This is a rather
general condition if we consider the finite length of the
roots. Therefore, our findings support the notion that, under
fairly broad conditions, only long-range competition is
required for patterns to occur, and suggests that the role of
short-range facilitation mechanisms may not be as fundamental
to pattern formation as has previously been thought.

\appendix

\section{Calculation of the perturbation growth rate}
\label{deriv} We start from Eq.~(\ref{model}) and perform a
complete linear stability analysis to obtain the
perturbation growth rate of Eq.~(\ref{condition}). The
objective of this technique, broadly used in the study of
nonlinear phenomena, is to obtain the temporal evolution of
small perturbations to the stationary homogeneous state of the
system. Considering small perturbations, the density is
$\rho({\bf x},t)=\rho_{0}+\epsilon\psi({\bf x},t)$, with
$\epsilon \ll 1$. Substituting it into the model equation
(\ref{model}), neglecting nonlinear terms in the perturbation,
and performing a first-order Taylor expansion of the
probability of overcoming competition, $r$, we obtain an
equation for the evolution of the perturbation

\begin{eqnarray}\label{pert_lin}
\frac{\partial \psi({\bf x},t)}{\partial t}&=&\beta_{0}r(\rho_{0},\delta)(1-2\rho_{0}) \psi({\bf x},t)-\alpha\rho_{0} \psi({\bf x},t)+ \nonumber \\
&&\beta_{0}r'(\rho_{0},\delta)\rho_{0}(1-\rho_{0})\int G(|{\bf x}-{\bf x'}|)\psi({\bf x'},t)d{\bf x'},\nonumber \\
\end{eqnarray}
which is a linear integro-differential equation with constant
coefficients that can be solved using the Fourier transform.
The transformed equation is

\begin{eqnarray}\label{transformed}
\frac{\partial \hat{\psi}({\bf k},t)}{\partial t}&=&\beta_{0}r(\rho_{0},\delta)(1-2\rho_{0}) \hat{\psi}({\bf k},t)-\alpha\rho_{0} \hat{\psi}({\bf k},t)+ \nonumber \\
&&\beta_{0}r'(\rho_{0},\delta)\rho_{0}(1-\rho_{0})\hat{G}({\bf k})\hat{\psi}({\bf k},t),
\end{eqnarray}
where $\hat{\psi}({\bf k},t)=\int{\rm e}^{i k\cdot x}\psi({\bf
x},t)d{\bf x}$ is the Fourier transform of the perturbation,
and equivalently, $\hat{G}(k)$ is the Fourier transform of the
kernel.

Finally, Eq.~(\ref{transformed}) is solved by $\hat{\psi}({\bf
k},t)\propto \exp(\lambda({\bf k})t)$, with the following
expression for the linear growth rate of the perturbation

\begin{equation}\label{reldis}
 \lambda({\bf k})=\beta_{0}\left[r(\rho_{0},\delta)(1-2\rho_{0})+(1-\rho_{0})\rho_{0}r'(\rho_{0},\delta)\hat{G}({\bf k})\right]-\alpha.
\end{equation}

Using the equation for the stationary solution,
Eq.~(\ref{homo}), and Eq.(\ref{cond2}) for the probability of
overcoming competition we arrive at the expression of
Eq.~(\ref{condition}).

\begin{acknowledgments}
R.M-G. is supported by the JAEPredoc program of CSIC.
 R.M-G., C.L. and E.H-G acknowledge support from FEDER and MICINN
(Spain) through Grants No. FIS2012-30634 INTENSE@COSYP and CTM2012-39025-C02-01 ESCOLA.
\end{acknowledgments}


%

\end{document}